\documentclass[prb]{revtex4}
\usepackage{graphicx}
\def\no{\noindent}
\def\bc{\begin{center}}
\def\ec{\end{center}}
\def\vs{\vskip0.5cm}
\def\beq{\begin{equation}}
\def\eeq{\end{equation}}

\def\d{\downarrow}
\def\u{\uparrow}

\begin{document}

\title{ Finite $E\otimes\beta$ Jahn-Teller Systems: \\
a Continued-Fraction Approach}

\author{K. Ziegler}
\email{Klaus.Ziegler@Physik.Uni-Augsburg.de}
\affiliation{
Institut f\"ur Physik, Universit\"at Ausgburg, D-86135 Augsburg, Germany
}


\begin{abstract}
A recursive method is developed to treat electrons coupled to phonons.
It is applied to small systems with $E\otimes\beta$ Jahn-Teller coupling.
Two cases are considered, a model with one electron and two orbitals on 
a single site (related to the Rabi Hamiltonian) and a model with two 
electrons on two sites. The corresponding Green's functions are
represented by rational functions. It is found that the spectra change
substantially when one phonon couples to the electron but are relatively
robust under an increasing number of phonons. 
\end{abstract}


\maketitle

\section{Introduction}

Small systems, like molecules or clusters of atoms, have attracted
much attention in recent years because of new experimental techniques
that provide detailed information of their spectral properties 
\cite{hahn01,zhitenev02}. There are two mechanisms that control the
physics of these small systems, one is the tunneling of electrons
between different orbitals and different atoms, the other is the
coupling between electrons and vibrational modes (phonons) of the
molecules or clusters. For the latter the Jahn-Teller coupling
scheme is relevant. Small Jahn-Teller systems can also be understood
as building blocks for lattice Jahn-Teller systems which play a crucial
role in solid-state physics, for instance, in the form of
transition metal oxids \cite{khomskii02}.

The main problem of treating electrons that couple to phonons is that
even for small systems with one or a few electrons the Hilbert space is 
infinite dimensional. This implies a complex spectrum with
level crossing and avoided level crossing \cite{bishop01}.
There are various treatments of small Jahn-Teller systems,
e.g. exact numerical diagonalization with truncated phonon spectrum
\cite{ranninger92,alexandrov94,weisse01}, Monte-Carlo simulations \cite{berger95} 
or variational methods \cite{trugman}. In this paper a systematic recursive 
procedure for treating the phonons in small electronic systems is developed and
applied to several examples. It is based on a projection formalism \cite{fulde}.
The method is quite flexible, can easily deal with degeneracies and was
previously introduced to two-component bosons on a lattice \cite{ziegler03}. 
The central idea is to approximate the elements of a Green's function 
systematically by standard (e.g. rational) functions. 

The paper is organized as follows: In Sect. 2 the evolution of a quantum
state and its connection to a projected Green's function is briefly
discussed. The continued-fraction approach is developed in Sect. 3. Then in 
Sect. 4 two models with Jahn-Teller coupling are introduced, a single-site
model where an electron tunnels between two orbitals and a model
where two electrons tunnel between two sites. The application of the
continued-fraction approach to these models is explained in Sect. 5
and the results are discussed in Sect. 6.

\section{Evolution of States: Projected Dynamics}

The evolution of a quantum state $|\Psi_t\rangle$ during the time
interval $[0,t]$ is given by
\[
|\Psi_t\rangle =e^{iHt}|\Psi_0\rangle,
\]
where the Hamiltonian $H$ is measured in units of $\hbar$. 
A Laplace transformation for a positive time $t$ gives for $Im z<0$ a 
resolvent:
\beq
\int_0^\infty e^{-izt}|\Psi_t\rangle dt
=\int_0^\infty e^{-izt}e^{iHt}dt |\Psi_0\rangle
=(z-H)^{-1}|\Psi_0\rangle.
\label{evol1}
\eeq
Suppose that the initial state $|\Psi_0\rangle$ is from a restricted Hilbert
space of low energy, the Green's function $(z-H)^{-1}$
acts on restricted Hilbert space, represented by the projector $P_0$ and 
the projected resolvent $(z-H)^{-1}P_0$.
To evaluate the probability for the system to return to the initial
state $|\Psi_0\rangle$, the following quantity must be evaluated:
\[
\langle\Psi_0|\Psi_t\rangle
=\int_{-\infty}^\infty e^{izt}\langle\Psi_0|(z-H)^{-1}|\Psi_0\rangle dz.
\]
If $E_j$ are the eigenvalues of $H$ we can write for Eq. (\ref{evol1})
\[
\langle\Psi_0|(z-H)^{-1}|\Psi_0\rangle
=\sum_j{|\langle E_j|\Psi_0\rangle|^2\over z-E_j}.
\]
The poles of this expression are the characteristic frequencies
of the evolution, starting from the projected Hilbert space and
returning to it. The imaginary part of the Green's function
gives the corresponding spectral density.  

\section{Projection Formalism and Continued-Fraction Representation}

After this preparation the goal is to evaluate the projected Green's function 
$P_0(z-H)^{-1}P_0$, where $P_0$ projects the states of the entire Hilbert 
space to the subspace ${\cal H}_0$.  It satisfies the identity
\beq
P_0(z-H)^{-1}P_0=\Big[
P_0(z-H)P_0 - P_0H P_1 (z-H)_1^{-1}P_1 HP_0\Big]_0^{-1},
\label{projected}
\eeq
where $P_1={\bf 1}-P_0$ projects onto the Hilbert space ${\cal H}_1$ that is
complementary to ${\cal H}_0$. If $H$ satisfies
\[
P_0HP_1=P_0HP_2,\ \  P_1HP_0=P_2HP_0\ \ \ (P_2\ne P_1),
\]
Eq. (\ref{projected}) becomes
\beq
P_0(z-H)^{-1}P_0=\Big[P_0(z-H)P_0 - P_0H P_2 (z-H)_1^{-1}P_2 HP_0\Big]_0^{-1}.
\label{projected2}
\eeq
The identity used in Eq. (\ref{projected}) can be applied again to
$P_2 (z-H)_1^{-1}P_2$ on the right-hand side. In general, a hierarchy of
projectors $P_k$ 
onto Hilbert spaces ${\cal H}_k$ is produced recursively. It is based 
on the fact that the projector $P_{2j+1}$ 
is created from $P_{2j-1}$ and $P_{2j}$ as
\[
{\cal H}_{2j+1}={\cal H}_{2j-1}\backslash {\cal H}_{2j}
\subset {\cal H}_{2j-1},
\]
and $P_{2j+2}$ comes from the relation
\[
P_{2j+1}HP_{2j}=P_{2j+2}HP_{2j}
\ \ \ {\rm and}\ \ \ 
P_{2j}HP_{2j+1}=P_{2j}HP_{2j+2}.
\]
This construction implies  
a recursion relation that connects pairs of projected Greens's functions:
\[
P_{2j}(z-H)_{2j-1}^{-1}P_{2j}=
\Big[P_{2j}(z-H)_{2j-1}P_{2j} 
- P_{2j}HP_{2j+2}(z-H)_{2j+1}^{-1}P_{2j+2}HP_{2j}\Big]_{2j}^{-1}.
\]
By using $G_{2j}=P_{2j}(z-H)_{2j-1}^{-1}P_{2j}$ and $H_{j,j+1}=P_{2j}HP_{2j+2}$
this reads
\beq
G_{2j}=\Big[z-P_{2j}HP_{2j} 
- H_{j,j+1}G_{2j+2}H_{j+1,j}\Big]_{2j}^{-1}.
\label{projected3}
\eeq
Further simplifications are possible if it is assumed that the Hamiltonian 
$H$ can be written as a sum of two Hamiltonians as $H=H_0+H_1$ with the 
following properties:

\no
(1) $H_0$ must stay inside the projected Hilbert space: $H_0P_0=P_0H_0P_0$
and $P_0H_0=P_0H_0P_0$.

\no
(2) $H_1$ maps from ${\cal H}_{2j}$ to ${\cal H}_{2j+2}$:
\[
H_1: {\cal H}_{2j}\to {\cal H}_{2j+2},
\]
where ${\cal H}_{2j}$ is orthogonal to ${\cal H}_{2j+2}$. Examples
shall be discussed subsequently.

\section{The Models: $E\otimes \beta$ Jahn-Teller coupling }

The Jahn-Teller electron-phonon coupling describes electrons that may
occupy two possible orbitals. The latter can be understood formally as
a pseudospin with values $\sigma=\u ,\d$.
On a molecule, a cluster or a lattice one phonon mode couples to 
electrons at each site, where the interaction enery is a sum over all 
sites $j=1,...$: 
\[
H_{eph}=g\sum_j(b_j^\dagger+b_j)(n_{j\u}-n_{j\d}).
\]
$n_{j\sigma}$ is the electronic number operator at site $j$ and orbital
$\sigma$, and $b_j^\dagger$ ($b_j$) is the creation (annihilation)
operator of a phonon.

The simplest case considered in this paper is a single site
with two orbitals, where an electron can tunnel between the two 
orbitals. This problem is also well-known in atomic physics under the name 
of Rabi Hamiltonian \cite{rabi37}.
An extension is a two-site system with two orbitals per site and with two 
electrons. The model is simple if both electrons are in the same orbital 
because of Pauli blocking. Therefore, we consider the case where the two 
electrons sit in different orbitals. They can tunnel between the two
sites and are coupled to each other via the electron-phonon interaction.

\subsection{Single site: Jahn-Teller effect with interorbital tunneling}

There are two electronic states, either the electron is in orbital $\u$
or in orbital $\d$. Moreover, there are
$N$ ($N=0,1,...$) phonons. Thus the Hilbert space is spanned by
\[
|N,\u\rangle,\ \  |N,\d\rangle .
\]
For electronic tunneling rate $t$ between the
two orbitals the resulting Rabi Hamiltonian reads
\[
H=t\sigma_1+\omega_0 \sigma_0 b^\dagger b+g\sigma_3(b^\dagger+b),
\]
where the Pauli matrices refer to the electronic states.

\subsection{Two sites: Jahn-Teller effect with intersite tunneling}

The extension of the model to two sites and two electrons provides
the opportunity to study the effect of electron-electron interaction
and electronic correlations. For a given pair of electrons, represented by 
integer numbers $n_{j \sigma}=0,1$, states with two electrons that occupy
different orbitals are considered. (Electrons stay in their orbital 
because inter-orbital tunneling is not included.) Then the following 
electronic states are available:
\beq
|\d,\u\rangle,\  |\u,\d\rangle,\  |\d\u,0\rangle,\  |0,\d\u\rangle.
\label{basis}
\eeq
The tunneling of the electrons and their Coulomb (Hubbard-like) interaction is 
defined by the Hamiltonian $H_e$ and by the dispersionless phonons with
energy $\omega_0$ as 
\[
H_e=-t\sum_{\sigma=\u,\d}(c_{1\sigma}^\dagger c_{2\sigma}
+c_{2\sigma}^\dagger c_{1\sigma})+U\sum_{j=1,2} n_{j\u}n_{j\d},\ \ \
H_{ph}=\omega_0\sum_{j=1,2}b_j^\dagger b_j .
\]
$c_j^\dagger$ ($c_j$) is the creation (annihilation) operator of a electron.
The electronic spin is not taken into accout here, i.e. spin-polarized
states are considered.
The electron-phonon interaction reads
\[
H_{eph}=g\sum_{j=1,2}(b_j^\dagger+b_j)(n_{j\u}-n_{j\d}),
\]
leading to the total Hamiltonian $H=H_e+H_{ph}+H_{eph}$.

\section{Application of the Continued-Fraction Approach to Small Systems}

If $H_1$ is a small perturbation to $H_0$ the resolvent of $H=H_0+H_1$
can be written as a Neumann series:
\[
(z-H)^{-1}=(z-H_0-H_1)^{-1}=(z-H_0)^{-1}\sum_{l\ge0}[H_1(z-H_0)^{-1}]^l.
\] 
Truncation after a finite number of terms yields poles only from the zeros 
of $z-H_0$. This is often insufficient
to observe a realistic pole structure of the Green's function $(z-H)^{-1}$. 
The approximation can be improved by using a Pad\'e approximation 
\cite{tremblay} or
a partial summation of infinitely-many contributions \cite{fulde}.
A systematic approach is the continued fraction of Sect. 3 that
approximates the projected Green's function $P_0(z-H)^{-1}P_0$ by rational functions
with a complex pole structure. Depending on the regime (weak or strong 
electron-phonon interaction) there are two different approximation
schemes.

\subsection{Strong Electron-Phonon Interaction}

If the tunneling energy is small in comparison with the
electron-phonon interaction it is possible to separate the Hamiltonian
as $H=H_0+H_1$ with
\beq
H_0=\omega_0\sigma_0 b^\dagger b+g\sigma_3(b^\dagger+b),\ \ \ H_1=t\sigma_1
\label{ham2}
\eeq
and consider $H_1$ as a perturbation in the sense of the discussion in
Sect. 3. To diagonalize $H_0$, the Lang-Firsov transformation can be used
as a unitary transformation
\[
u=\pmatrix{
e^{\alpha(b^\dagger -b)} & 0 \cr
0 & e^{-\alpha(b^\dagger -b)}\cr
}, \ \ \ \alpha=g/\omega_0
\]
with
\[
u H_0u^\dagger =H_0'=\omega_0 (b^\dagger b -\alpha^2)\sigma_0
\]
The transformation of $H_1$ creates a complicated expression $H_1'$
that connects states with different phonon numbers. This makes it difficult 
to perform the iteration of the recursion relation.
For $P_0$ being the projection on the Hilbert space with $N=0$ phonons 
{\it after} the Lang-Firsov transformation was applied, 
Eq. (\ref{projected}) gives
\[
P_0(z-H')^{-1}P_0
=\Big[
P_0(z-H_0')P_0 - P_0H_1' P_1 (z-H')_1^{-1}P_1 H_1'P_0\Big]_0^{-1},
\]
and by using the approximation $(z-H')_1^{-1}\approx (z-H_0')_1^{-1}$ one
obtains
\[
P_0(z-H')^{-1}P_0\approx
\Big[
P_0(z-H_0')P_0 - P_0H_1' P_1 (z-H_0')_1^{-1}P_1 H_1'P_0\Big]_0^{-1}.
\]
After a lengthy but straightforward calculation this becomes
\beq
=\Big[z-{t^2e^{-\alpha^2}\over \omega_0}\Big({1\over(z/\omega_0+\alpha)}
-\gamma^*(-z/\omega_0-\alpha^2,-4\alpha^2)\Big)\Big]^{-1}\sigma_0
\label{scgf}
\eeq
with the incomplete Gamma function \cite{abramowitz}
\[
\gamma^*(a,y)=\sum_{m\ge0}{1\over m!}{(-y)^m\over a+m}.
\]
The renormalization factor $e^{-\alpha^2/2}$ of the tunneling rate $t$
is a well-known effect of the phonons, originally established in polaron
physics \cite{ranninger92,alexandrov94}, and also observed in the 
strong-coupling regime of the Hubbard-Holstein model \cite{sawatzky04}. 
Results of the iteration are shown in Fig. 1 for the spectral density.
It should be noticed that the expression for $P_0(z-H)^{-1}P_0$
with the projection $P_0$ {\it before} the Lang-Firsov transformation 
was applied is more complicated.

\subsection{Weak Electron-Phonon Interaction}

If the tunneling energy is large in comparison with the
electron-phonon interaction a different separation of the Hamiltonian
$H=H_0+H_1$ is needed:
\[
H_0=t\sigma_1+\omega_0 b^\dagger b\sigma_0,\ \ \ H_1=g\sigma_3(b^\dagger+b).
\]
Eq. (\ref{projected3}) can be truncated for $N$ phonons. This leads to the 
equations
\[
G_{2j}=\Big[
z\sigma_0-t\sigma_1-j\sigma_0-g^2\sigma_3 b G_{2j+2}b^\dagger\sigma_3
\Big]^{-1}_{N-j}
\]
with terminating condition
\[
G_{2N}=((z-N)\sigma_0-t\sigma_1)^{-1}
={1\over (z-N)^2-t^2}\pmatrix{
z-N & t \cr
t & z-N \cr
}.
\]
Some results of the iteration for the spectral density are shown in Fig. 2. 

\subsection{Two sites with two electrons} 

The Hamiltonian reads in the basis of Eq. (\ref{basis})
\beq
H_e=\pmatrix{
0 & 0 & -t & -t \cr
0 & 0 & t & t \cr
-t & t & U & 0 \cr
-t & t & 0 & U \cr
},\ \ 
H_{ph}=\omega_0(b_1^\dagger b_1+b_2^\dagger b_2)\pmatrix{
1 & 0 & 0 & 0 \cr
0 & 1 & 0 & 0 \cr
0 & 0 & 1 & 0 \cr
0 & 0 & 0 & 1 \cr
},
\label{ham3}
\eeq
and
\beq
H_{eph}=g\pmatrix{
-\Delta^\dagger-\Delta & 0 & 0 & 0 \cr
0 & \Delta^\dagger+\Delta & 0 & 0 \cr
0 & 0 & 0 & 0 \cr
0 & 0 & 0 & 0 \cr
}\equiv g(\Delta^\dagger+\Delta)S,
\label{ham4}
\eeq
where $\Delta=b_1-b_2$. Thus $\Delta$ ($\Delta^\dagger$) lowers (raises) the
number of phonons by one.

Using $H_0=H_e+H_{ph}$ and $H_1=H_{eph}$ (this is weak-coupling case)
the recursion relation of Eq. (\ref{projected3}) reads
\beq
G_{j}=\Big[z-P_j(H_e+H_{ph})P_j-g^2P_j\Delta S
G_{j+1}S\Delta^\dagger P_j\Big]^{-1}_j
\label{rec3}
\eeq
with the terminating condition for a maximum of $N$ phonons:
\beq
G_N=\Big[z-P_N(H_e+H_{ph})P_N\Big]^{-1}_N.
\label{rec4}
\eeq
This $4(N+1)\times 4(N+1)$ matrix is diagonal in terms of the 
phonon states. There are $N+1$ different phonon states, since there can
be $k$ $(=0,1...,N)$ phonons at the first site and $N-k$ phonons at the
second site. 

The action of the phonon operators $P_j\Delta$ and $\Delta^\dagger P_j$ on 
the $4(j+2)\times 4(j+2)$ phonon-diagonal matrix $G_{j+1}$
creates a $4(j+1)\times 4(j+1)$ phonon-diagonal matrix matrix with $4\times 4$ 
matrices $D(k,j+1-k)=G_{j+1}(k,j+1-k|k,j+1-k)$:
\[
P_j\Delta\pmatrix{
D(0,j+1) & 0 & ... & 0 \cr
0 & \ddots & \ddots & \vdots \cr
\vdots & \ddots & \ddots & 0 \cr
0 & ... & 0 & D(j+1,0)\cr
}\Delta^\dagger P_j
\]
\[
=2\pmatrix{
D(0,j) & 0 & ... & 0 \cr
0 & \ddots & \ddots & \vdots \cr
\vdots & \ddots & \ddots & 0 \cr
0 & ... & 0 & D(j,0)\cr
}
\]
This can be used to perform the iterations according to Eqs. (\ref{rec3})
and (\ref{rec4}). The results for the spectral density of the electronic states 
$|\u\,d\rangle$ and $|\u\d,0\rangle$ with a maximum of $N=3$ phonons are 
shown in Fig. 3. 

\section{Discussion}

The recursive evaluation of the projected Green's function of Sects. 5.2 and
5.3 can be
easily performed with an algebraic manipulation program. To demonstrate
the qualitative tendencies the case for small numbers of phonons $N=1,2,3$
is discussed in the following. It should be noticed that much higher
numbers can be studied with little effort. They show the same tendencies
as those of small $N$.

\subsection{Single-site model}

In the strong-coupling regime the tunneling between the orbitals can be 
completely suppressed (i.e., $t=0$). Then the projected Green's function has 
only one pole, namely $z=0$. For any $t>0$, however, there are infinitely many
poles due to the incomplete Gamma function in Eq. (\ref{scgf}). But not all
poles contribute with the same weight, as it is shown in Fig. 1. There
are two effects: (i) Phonons lower the groundstate energy and 
(ii) the weight of the excitations decreases rapidly with increasing energy.

Eigenvalues in the weak-coupling regime in the absence of phonons (i.e. $N=0$,
where $H_0=t\sigma_1+\omega_0 b^\dagger b\sigma_0$) are $E=\pm t$. 
This is a level splitting caused by the tunneling between
the two orbitals. Already a single phonon lowers the groundstate and creates
new excited states, as shown in the second plot of Fig.2. Additional 
phonons do not affect the groundstate but shift excited states and create 
new ones. In Fig. 2 this is plotted for phonon numbers up to $N=3$.
The spectral weights of the excited states are also affected by the increasing
number of phonons. All these effects are related to fact that the elements of the
projected Green's function are rational functions
\beq
{P_N(z)\over Q_N(z)},
\label{pgf}
\eeq
where $N$ is the maximal number of phonons taken into account in the
virtual processes. For $t=g=1$ and $N=0,1,2,3$ the Green's function
$(\langle\u |(z-H)^{-1}|\u\rangle$ has been calculated with MAPLE as

\no  
$N=0$:
\[
{\frac {z}{{z}^{2}-1}}
\]
$N=1$:
\[
-{\frac {-4\,z+{z}^{3}+7\,{z}^{2}+{z}^{5}-4\,{z}^{4}-1}{-5+4\,{z}^{2}
+12\,z+{z}^{4}-14\,{z}^{3}+4\,{z}^{5}-{z}^{6}}}
\]
$N=2$:
\[
-{\frac {4+3\,z-46\,{z}^{2}+33\,{z}^{4}+24\,{z}^{3}+10\,{z}^{6}
-33\,{z}^{5}-{z}^{7}}{-11-64\,z+46\,{z}^{2}+94\,{z}^{3}
+31\,{z}^{6}-14\,{z}^{5}-82\,{z}^{4}-10\,{z}^{7}+{z}^{8}}}
\]
$N=3$:
\[
-{\frac {8-218\,z+701\,{z}^{2}-558\,{z}^{3}-449\,{z}^{6}+759\,{z}^{5}
-361\,{z}^{4}+128\,{z}^{7}+{z}^{9}-18\,{z}^{8}}{-180+872\,z-716\,{z}^{2}
-1156\,{z}^{3}-519\,{z}^{6}-440\,{z}^{5}+1809\,{z}^{4}+414\,{z}^{7}
-{z}^{10}+18\,{z}^{9}-126\,{z}^{8}}}
\]
Apparently, the order of the polynomial $P_N(z)$ is one less than the
order of $Q_N(z)$ and increases by 2 for $N\ge1$:
\[
P_N(z)=z^{2(N+1)+1}+o(z^{2(N+1)}),\ \ \ Q_N(z)=z^{2(N+2)}+o(z^{2(N+2)-1})
\]

\subsection{Two-site model}

In the case of two electrons on two sites there is tunneling between
the two sites. Without phonons the Hamiltonian $H_0=H_e$ in Eq. (\ref{ham3}) 
has eigenvalues
\[
E=0,\ \ U,\ \ {U\pm\sqrt{U^2+16t^2}\over2}
\]
Without Hubbard repulsion $U$ there are three different eigenvalues, where
the degeneracy of $E=0$ represents the states
\[
E=0: |\d,\u\rangle+|\d,\u\rangle,\ \ \ E=U: |\d\u,0\rangle+|0,\d\u\rangle.
\] 
With Hubbard repulsion $U>0$ there are four non-degenerate eigenvalues. 
The most obvious effect in this two-site model is the lowering of 
the groundstate energy already by a single phonon and the creation of
excitations on energies above the groundstate energy. It is interesting
to notice that more excitations contribute to the singly-occupied
Green's function $\langle\d,\u |(z-H)^{-1}|\d,\u \rangle$ than to the
doubly-occupied Green's function 
$\langle\d\u,0 |(z-H)^{-1}|\d\u,0 \rangle$.
Moreover, there is always a state with maximal weight at $E=U$ for 
the state $|\d\u,0\rangle$. 

The Green's function with maximally $N$ phonons is again a rational function
of the form (\ref{pgf}). For $t=g=1$ and $U=0$ the expression
\[
\langle\d,\u |(z-H)^{-1}|\d,\u \rangle={P_N(z)\over Q_N(z)}
\]
has been calculated with MAPLE as

\no
$N=0$:
\[
P_0(z)= z^2-2,\ \ \ Q_0(z)=z(z^2-4)
\]
$N=1$:
\[
P_1(z)=
(z^5-3z^4-7z^3+15z^2+8z-10)(z^2-2z-5)
\]
\[
Q_1(z)=z^8-5z^7-10z^6+62z^5+33z^4-221z^3-44z^2+176z-40
\]
$N=2$:
\[
P_2(z)=z^{10}-13z^9+49z^8+11z^7-378z^6+386z^5+720z^4-852z^3-536z^2+408z+144
\]
\[
Q_2(z)=z^{11}-13z^{10}+45z^9+61z^8-562z^7+396z^6+1788z^5-2092z^4
-1368z^3+1632z^2+224z-192
\]
$N=3$:
\[
P_3(z)=z^{13}-24z^{12}+226z^{11}-1000z^{10}+1578z^9+3192z^8-15756z^7
+15484z^6+16149z^5-35976z^4+8226z^3
\]
\[
+15076z^2-7944z+672
\]
\[
Q_3(z)=
z(z^{13}-24z^{12}+222z^{11}-906z^{10}+708z^9+7046z^8-22638z^7
+9330z^6+61467z^5-98222z^4+18512z^3
\]
\[
+56152z^2-38816z+7168)
\]
Apparently, the order of the polynomial $P_N(z)$ is one less than the
order of $Q_N(z)$ and increases by 3 for $N\ge1$:
\[
P_N(z)=z^{3(N+1)+1}+o(z^{3(N+1)}),\ \ \ Q_N(z)=z^{3(N+1)+2}+o(z^{3(N+1)+1}).
\]

\section{Conclusions}

A a recursive method has been developed to study the spectral properties of
small Jahn-Teller systems. It is based on a decomposition of the 
infinite-dimensional Hilbert space, spanned by a few electronic and
an unlimited number of phononic states. Two cases were considered, one for
an infinite number of phonons (the strong-coupling case) and one in which
the number of phonons is increased by one in each iteration step of a recursive
equation (the weak-coupling case). In both cases the iteration of the
recursion relation leads to a continued-fraction representation of a
projected Green's function. The matrix elements of the resulting Green's
function are related to the incomplete Gamma function in the strong-coupling
case and to rational functions in the weak-coupling case, respectively.

An advantage of this method is that it approaches the exact solution
systematically by standard functions. It is an alternative to perturbative
approaches, based on a power series of a model parameter (e.g. the tunneling 
rate $t$ or the electron-phonon coupling constant $g$). 
  
\vs

\no
Acknowledgement:

\no
I would like to thank K.-H. H\"ock and K. Becker for useful discussions.

\noindent
This work was supported by the Deutsche Forschungsgemeinschaft through
Sonderforschungsbereich 484.


\begin{figure} 
\begin{center}
\includegraphics[scale=0.6]{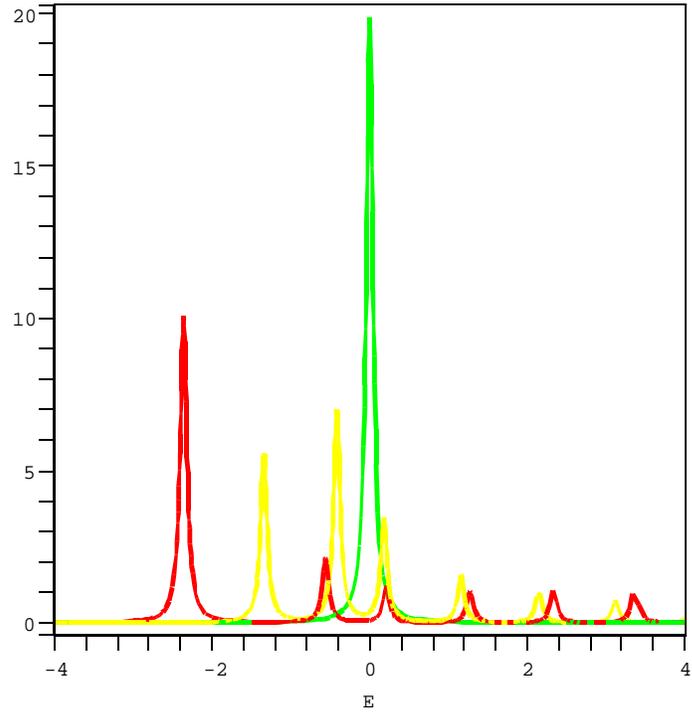}
\end{center} 
\caption{Strong coupling approach: The spectral density for $z=E+0.05i$ and
$g=1$, $N=1$, and $t=0, 0.5, 1$.}
\end{figure}

\begin{figure} 
\begin{center}
\includegraphics[scale=0.6]{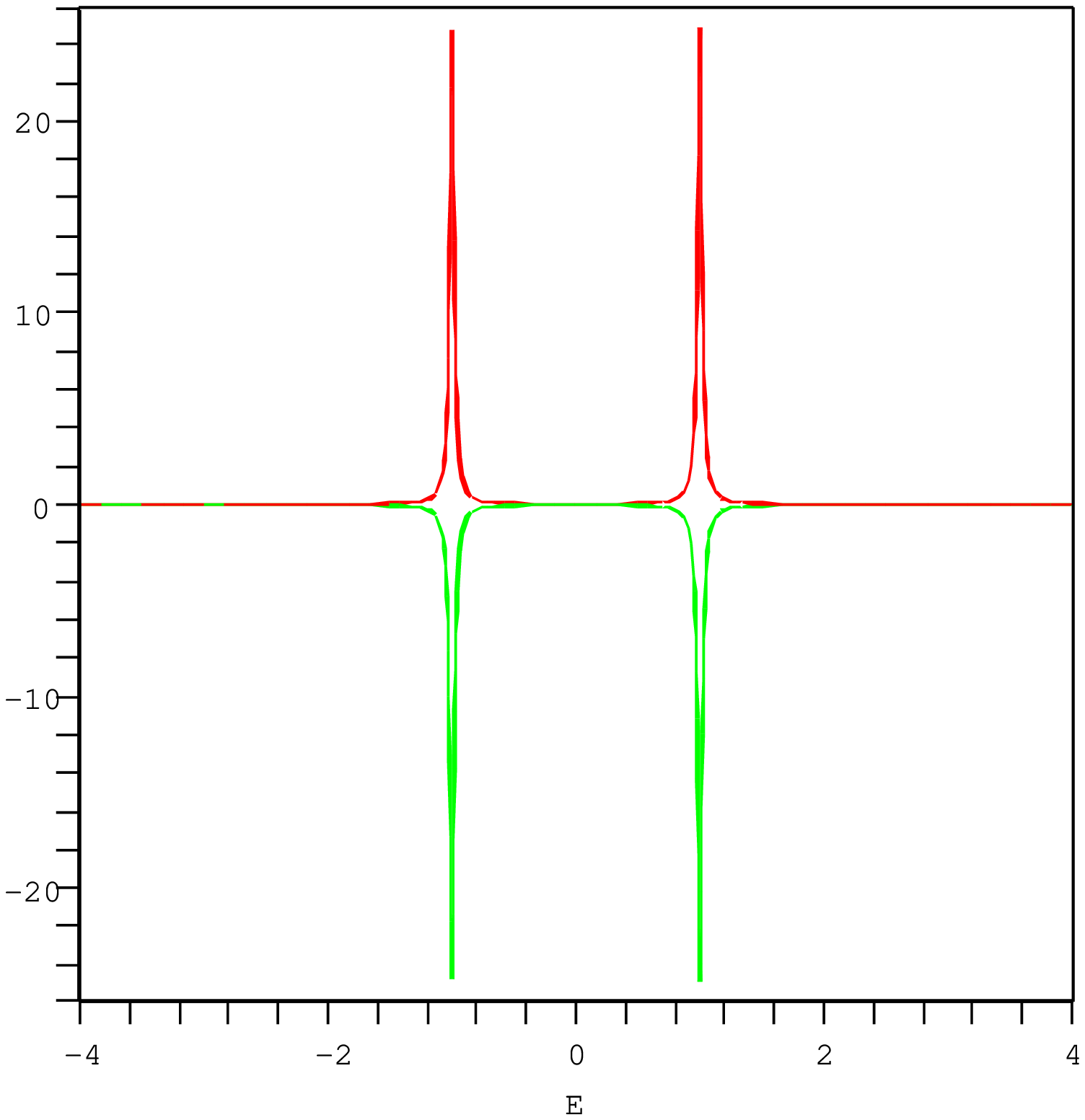}
\includegraphics[scale=0.6]{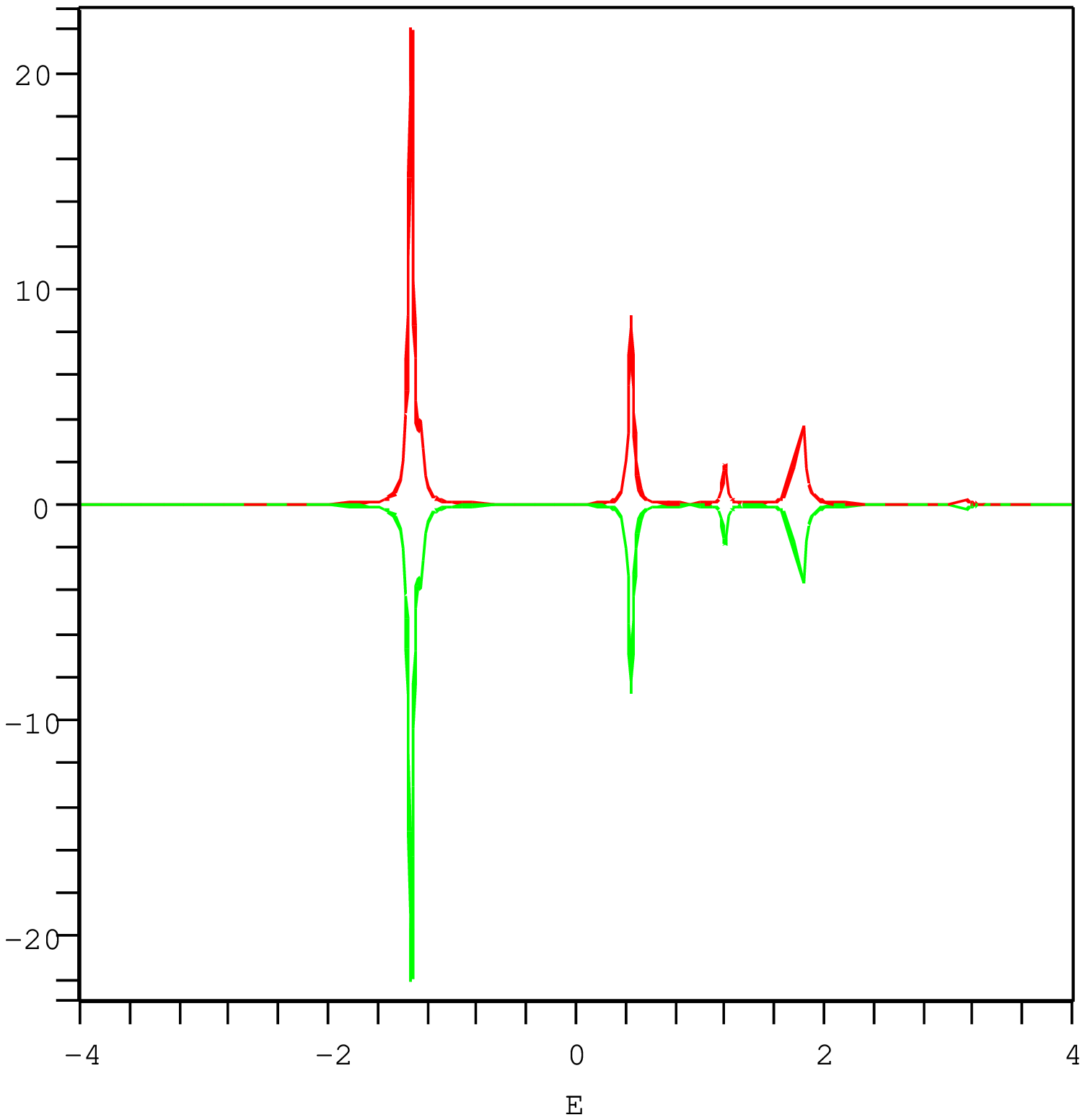}
\end{center} 
\caption{The spectral density of the projected Greens's function for
the single-site model with $t=g=1$ and $N=0,1$: 
$-Im(\langle\u |(z-H)^{-1}|\u\rangle$ (upper curves) and
$Im(\langle\d |(z-H)^{-1}|\d\rangle$ (lower curves).}
\end{figure}
\begin{figure} 
\begin{center}
\includegraphics[scale=0.6]{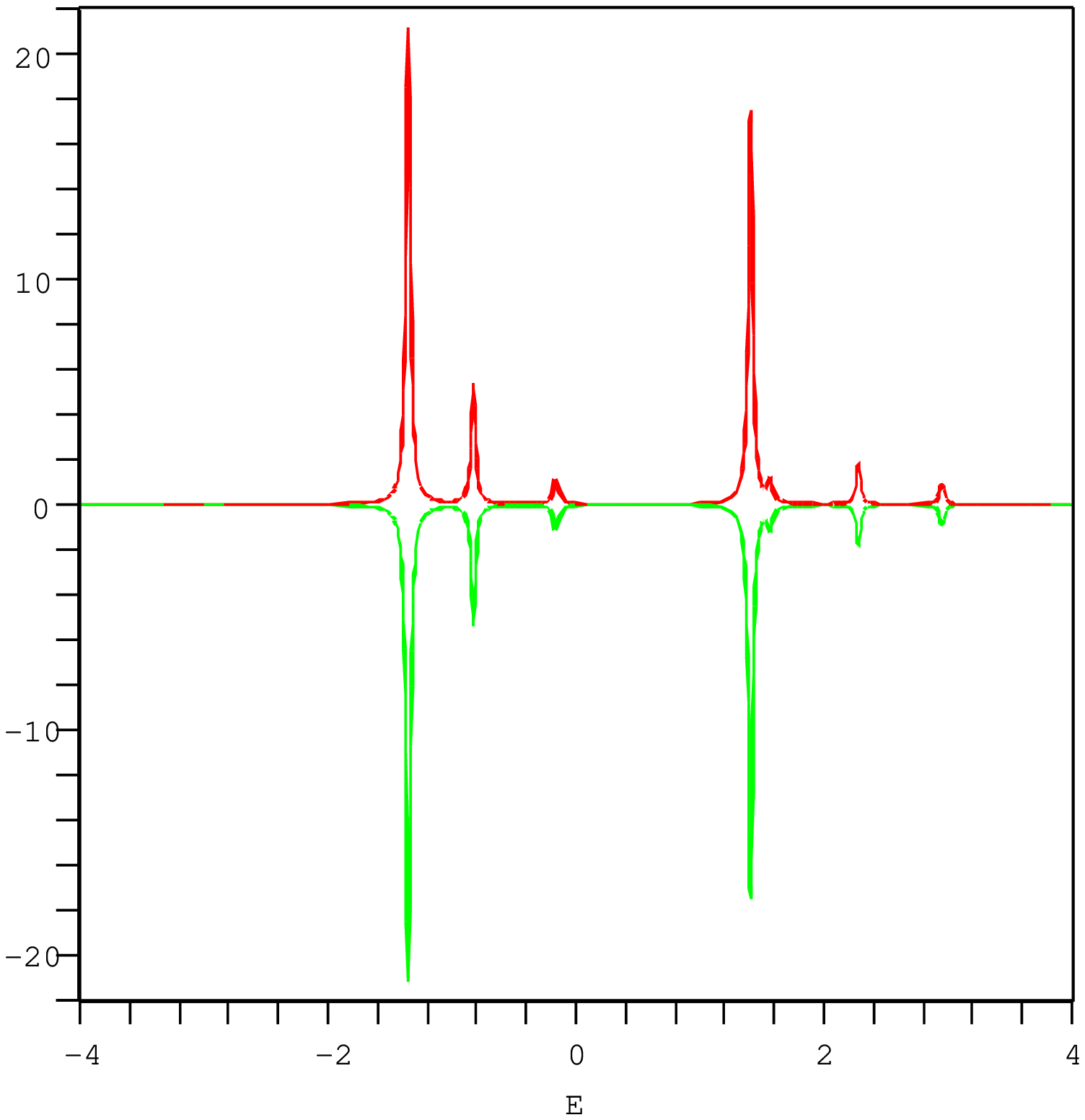}
\includegraphics[scale=0.6]{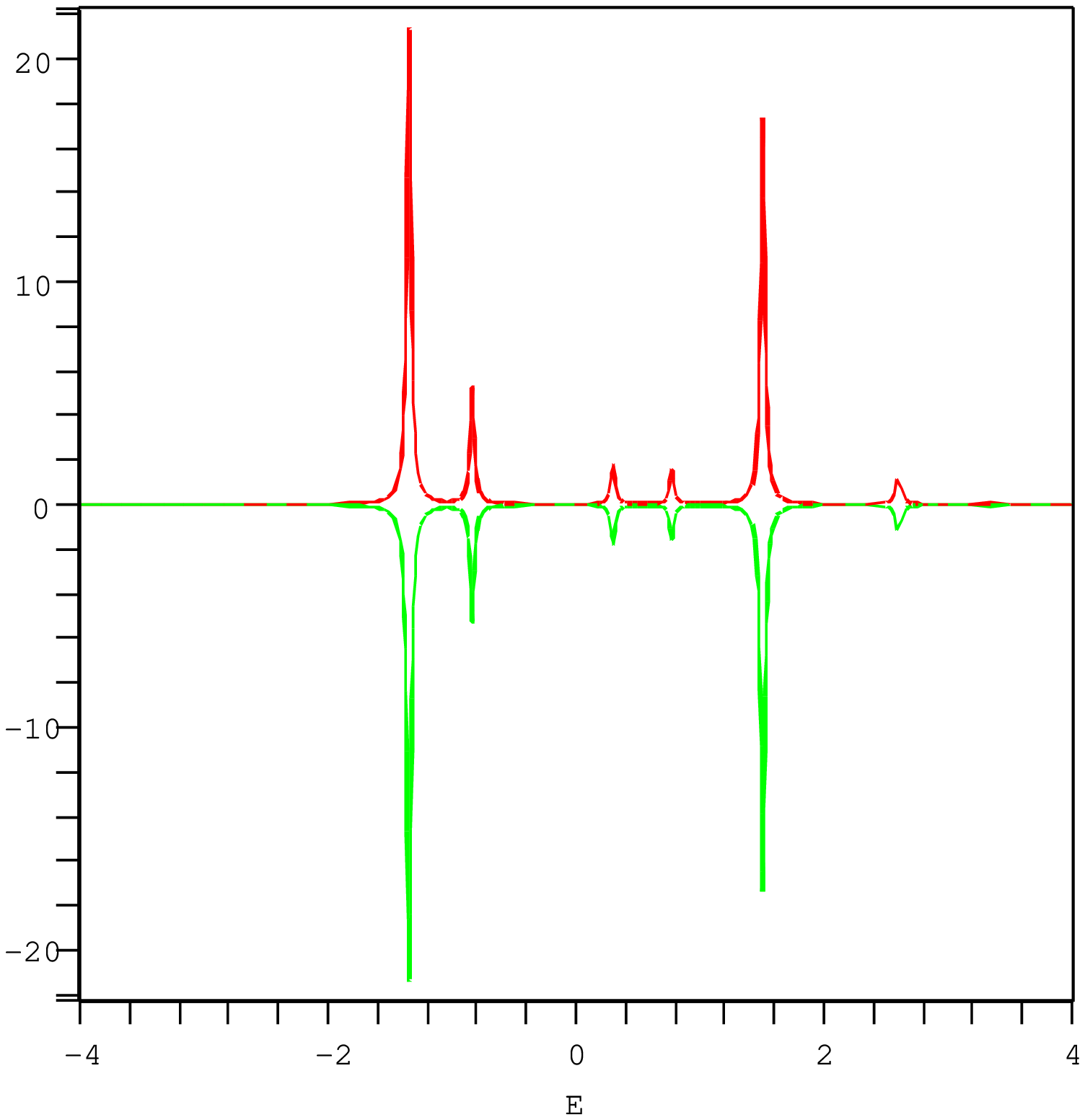}
\end{center} 
\caption{as previous Fig. but for $N=2,3$.}
\end{figure}

\begin{figure} 
\begin{center}
\includegraphics[scale=0.6]{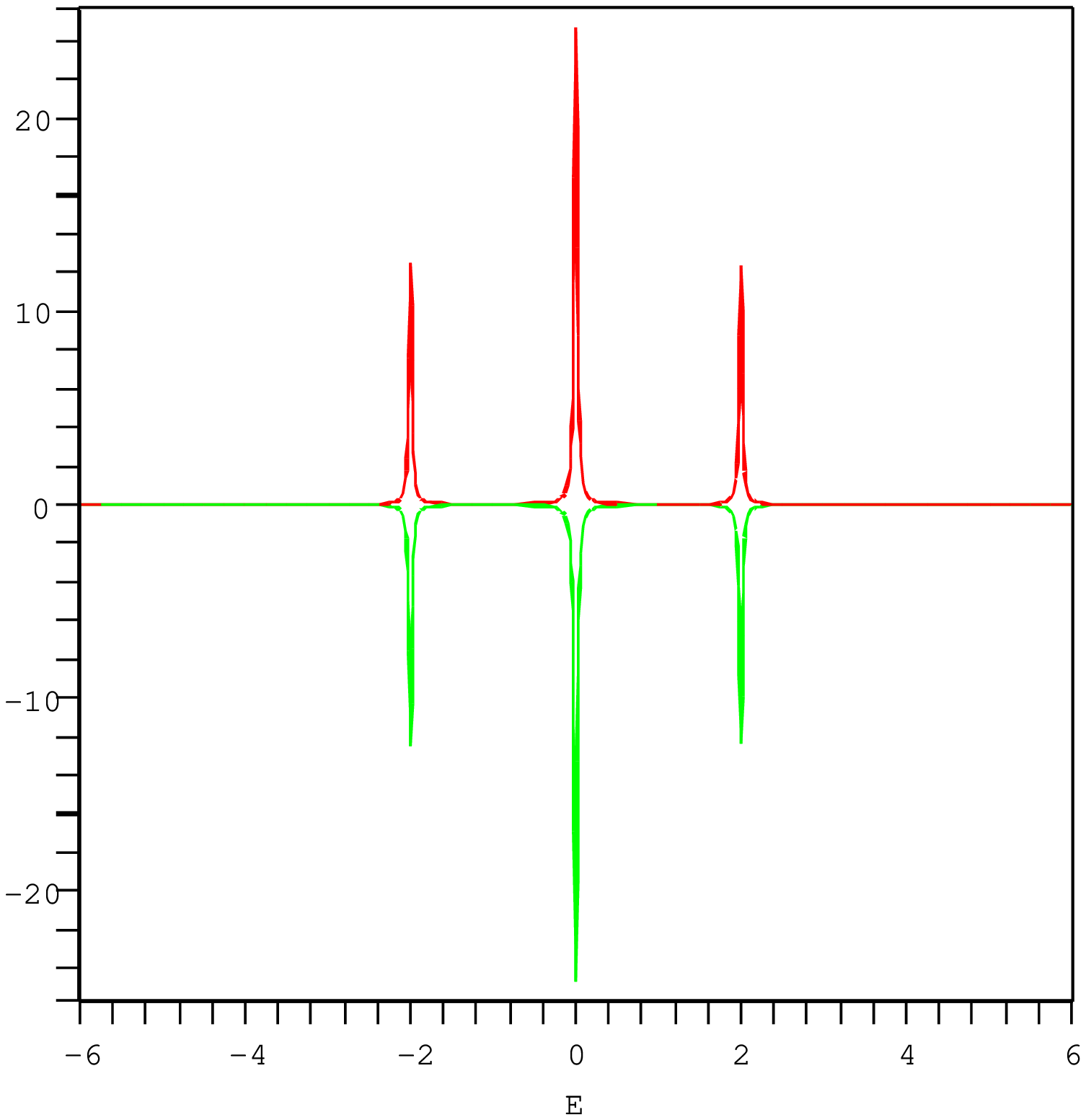}
\includegraphics[scale=0.6]{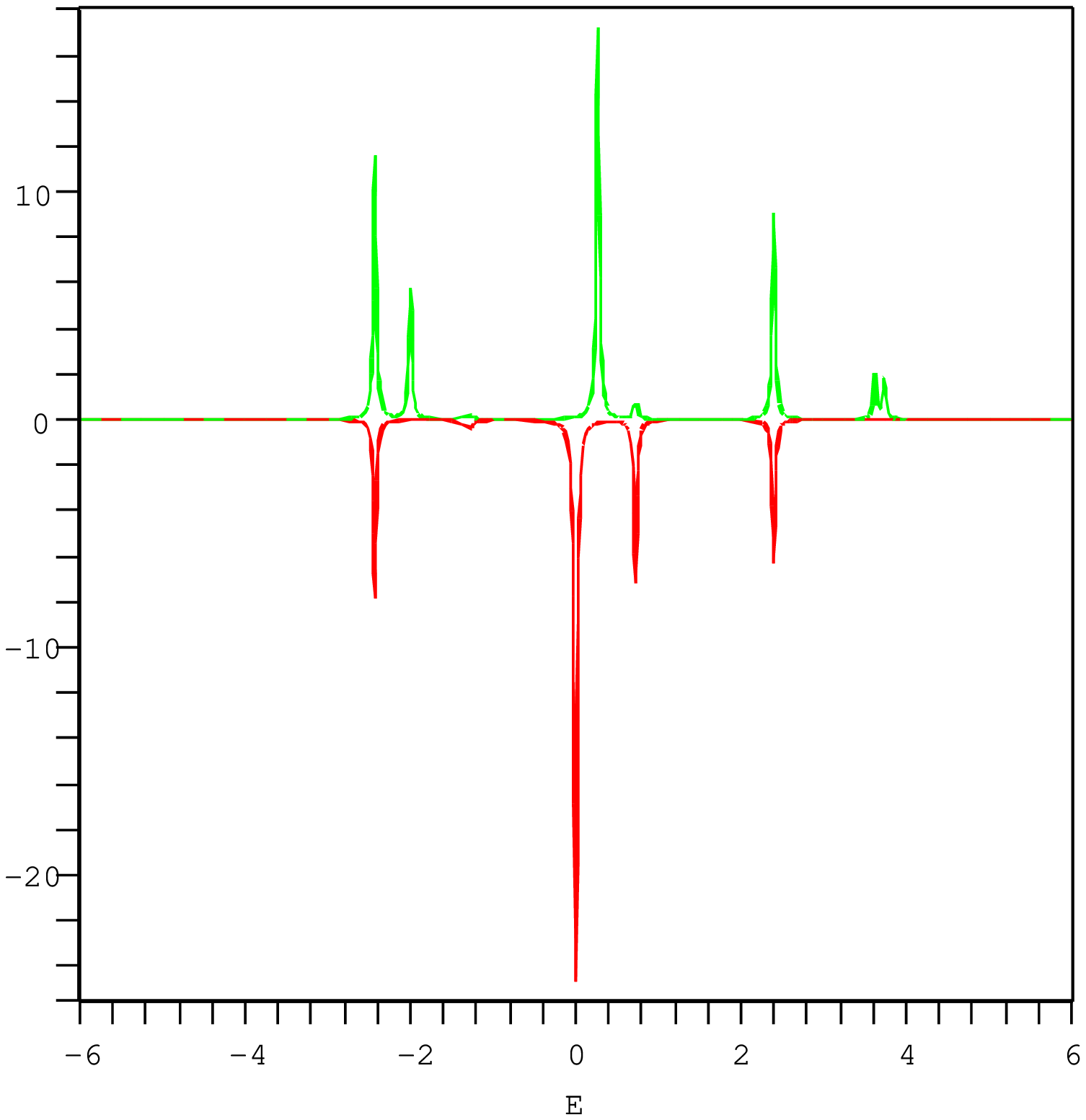}
\end{center} 
\caption{The spectral density of the projected Greens's function for
the two-site model with $t=g=1$, $U=0$, and $N=0,1$:
$-Im(\langle\d,\u |(z-H)^{-1}|\d,\u \rangle$ (upper curves) and
$Im(\langle\d\u,0 |(z-H)^{-1}|\d\u,0\rangle$ (lower curves).}
\end{figure}

\begin{figure} 
\begin{center}
\includegraphics[scale=0.6]{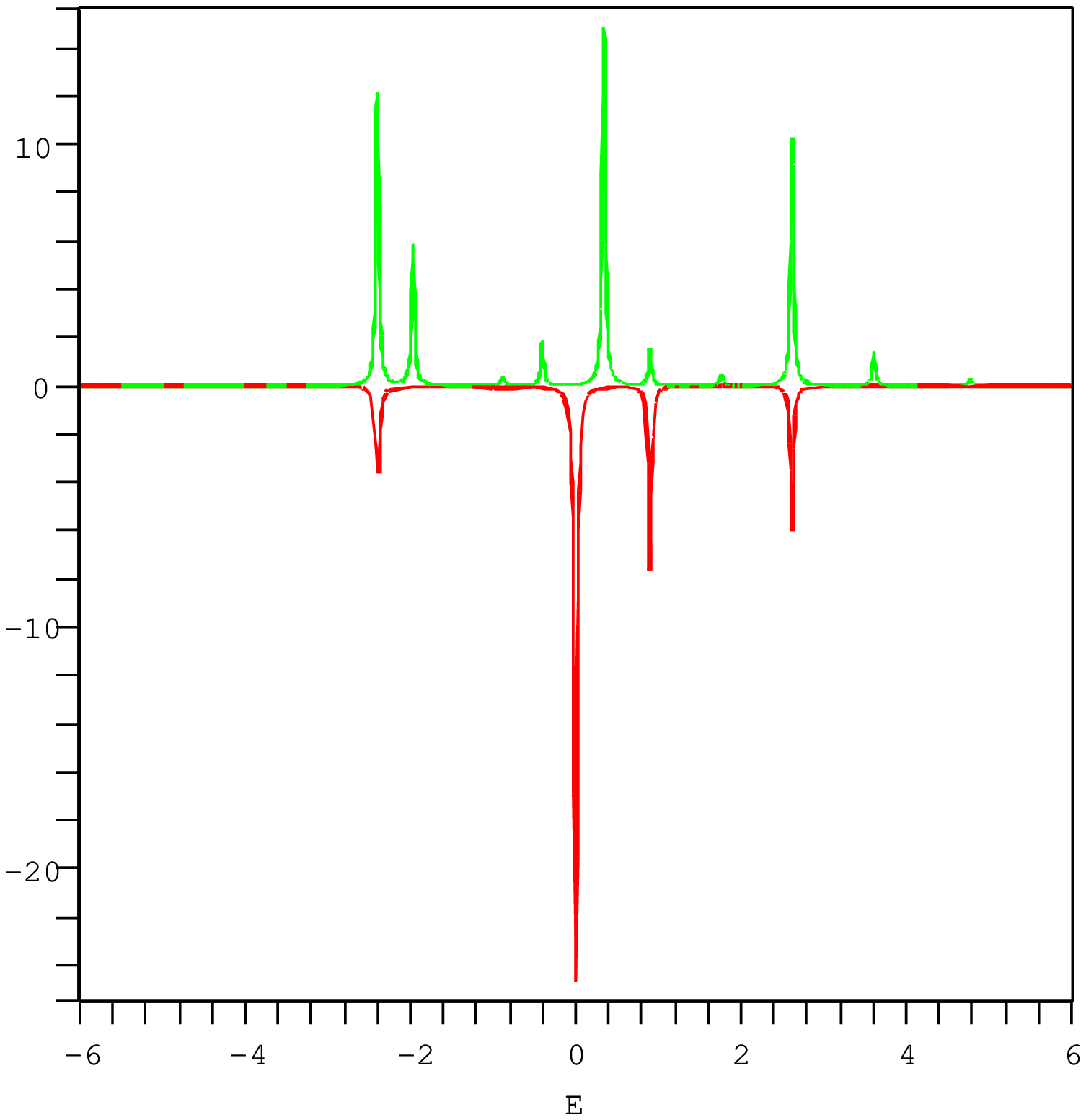}
\includegraphics[scale=0.6]{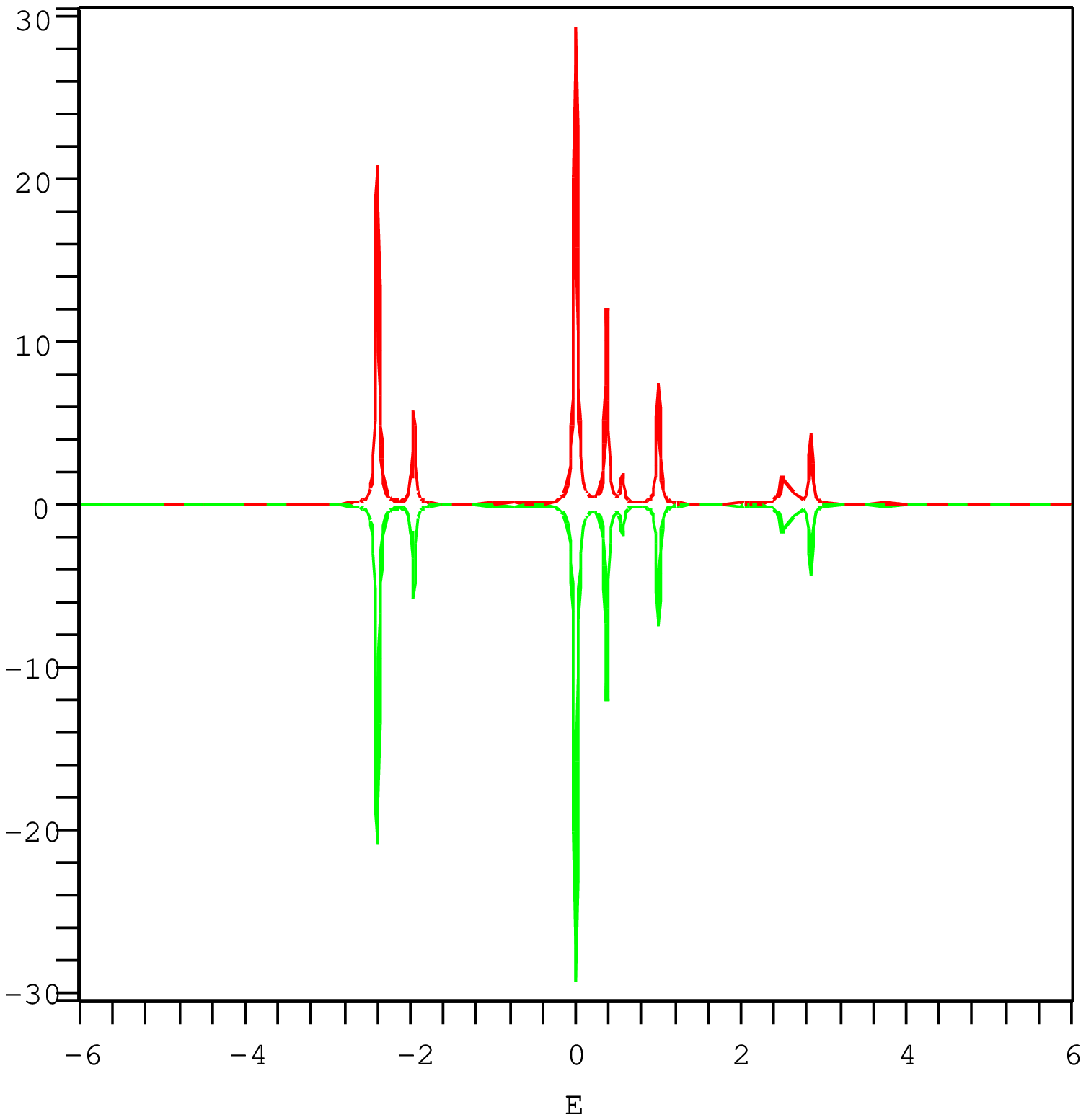}
\end{center} 
\caption{as previous Fig. but for $N=2,3$.}
\end{figure}

\begin{figure} 
\begin{center}
\includegraphics[scale=0.6]{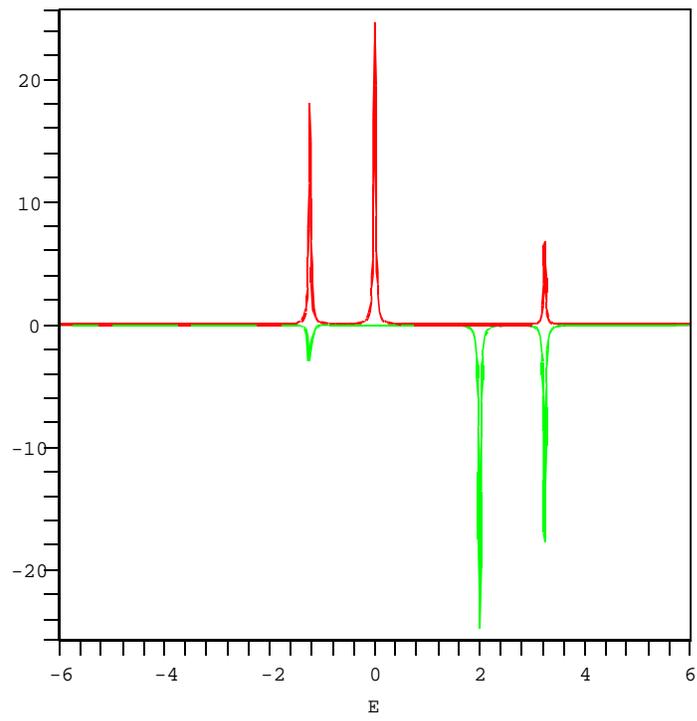}
\includegraphics[scale=0.6]{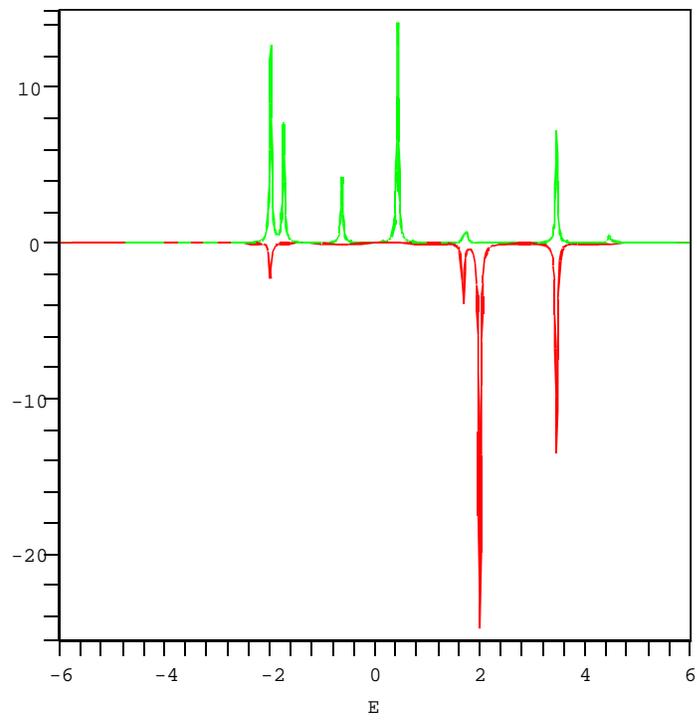}
\end{center} 
\caption{as previous Fig. but for $U=2$, $N=0,1$.}
\end{figure}

\begin{figure} 
\begin{center}
\includegraphics[scale=0.6]{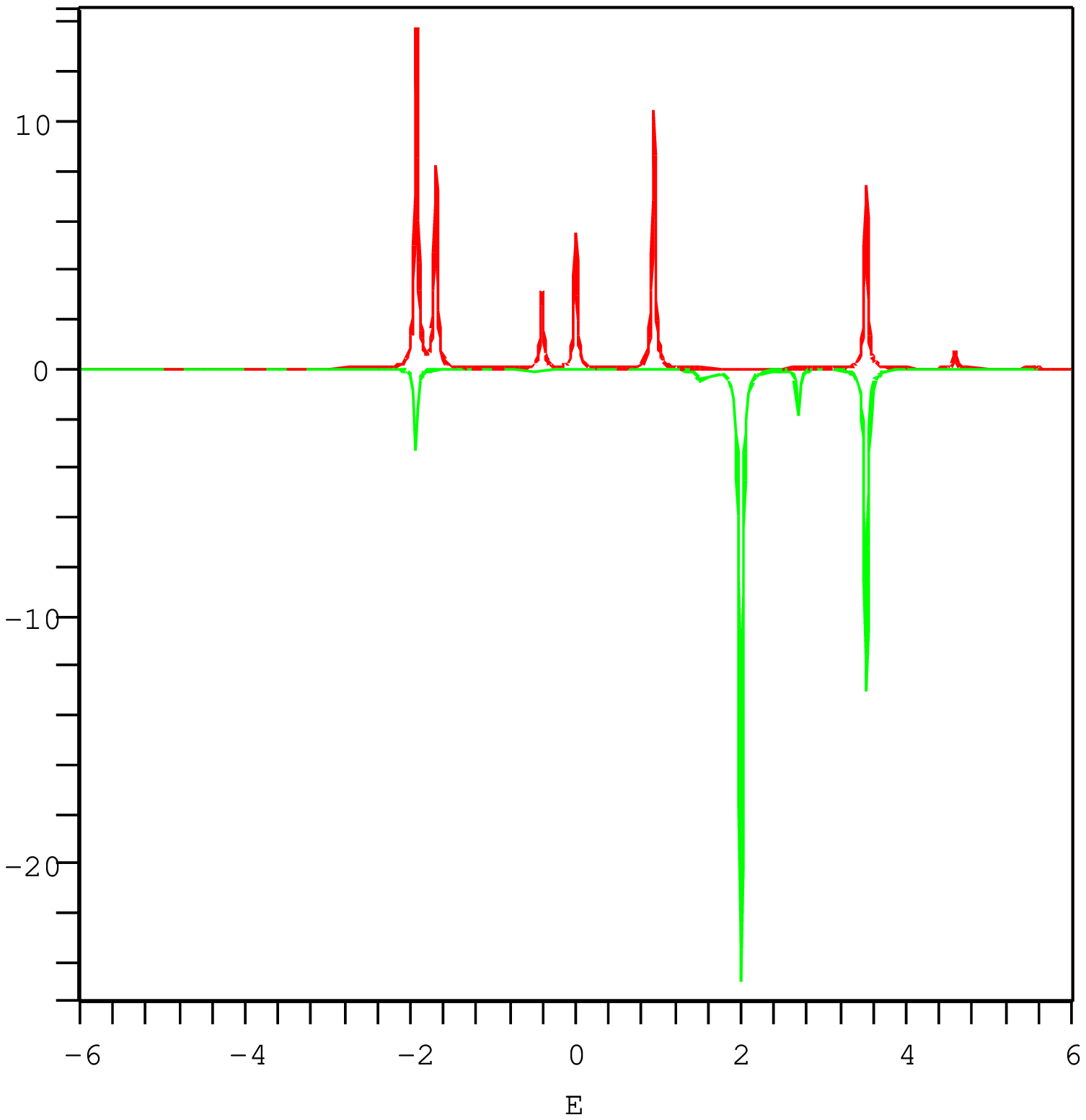}
\includegraphics[scale=0.6]{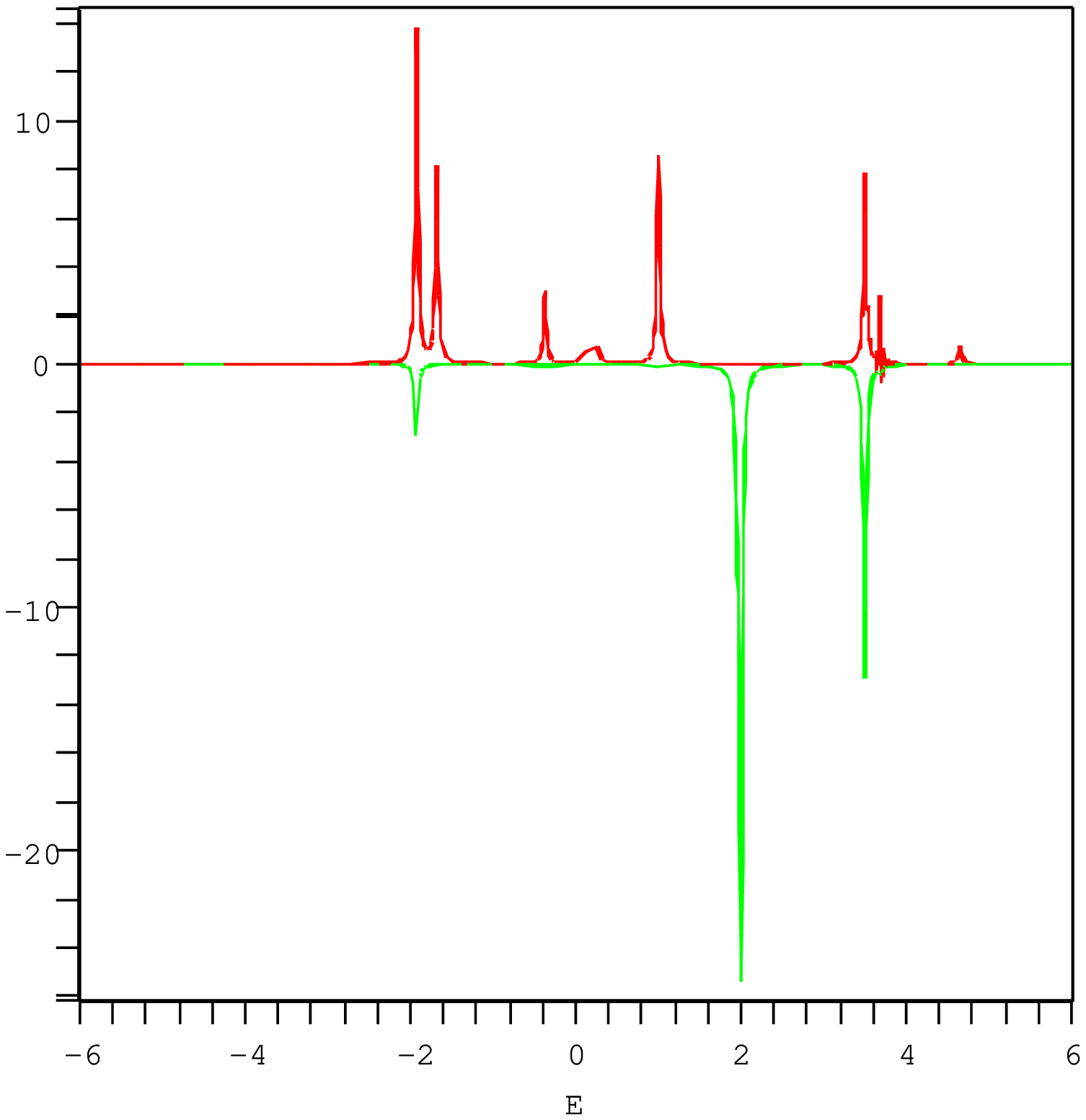}
\end{center} 
\caption{as previous Fig. but for $N=2,3$.}
\end{figure}

\end{document}